\documentclass[aps,pra,twocolumn,superscriptaddress]{revtex4}
\usepackage{graphicx}
\usepackage{amsmath,amssymb}
\begin{document}

\title{Readout of a weakly coupled qubit through the use of an auxiliary mode}

\author{Filippo Troiani}

\affiliation{Centro S3, CNR-Istituto di Nanoscienze, 
I-41125 Modena, Italy}

\date{\today}                                           


\begin{abstract} 
The dispersive coupling between a qubit and a cavity mode is widely used for performing non-destructive readout of the qubit state. In this approach, it is typically required that the dispersive strong coupling regime is achieved. Here we show that the use of an auxiliary cavity mode reduces by orders of magnitude the required value of the dispersive coupling, for a given decay rate of the cavity mode. The analysis is performed within the input-output formalism, in terms of the photon scattering matrix elements and of the signal-to-noise ratio. We derive simple analytical expressions for the optimal parameters and recover the standard single-mode result as a limiting case. The present results can also be applied to the qubit readout based on longitudinal cavity-qubit interactions, and to any sensing scheme where the cavity frequency is used as a probe to estimate some physical parameter of interest.
\end{abstract}

\maketitle

\section{Introduction}\label{SecI}

Cavities can be used for performing nondestructive measurements of qubit systems. In the presence of a transverse (Jaynes-Cummings) qubit-cavity interaction, the non-destructive character of the measurement can be approached in the dispersive regime, where the cavity is sufficiently detuned from the qubit to prevent an exchange of excitations between the two, while virtual transitions induce a qubit-state dependent shift of the resonance frequency. This results in a phase modulation of the (reflected or transmitted) field, which is typically detected in a homodyne measurement. Such an approach has been implemented on a variety of solid-state platforms, including superconducting qubits \cite{Blais04,Wallraff05,Filipp09,Vijay12,Hatridge13,Gu17} and semiconductor quantum dots \cite{Frey12,Petersson12}. Some of the limitations related to the use of the dispersive regime can be overcome in the presence of a longitudinal qubit-cavity interaction, which allows a faster read out of the qubit \cite{Didier15,Richer16}.

\begin{figure}
\begin{center}
\includegraphics[width=0.40\textwidth]{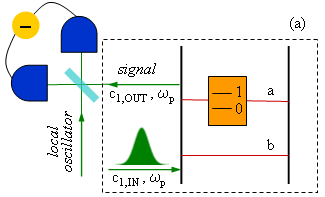}
\includegraphics[width=0.40\textwidth]{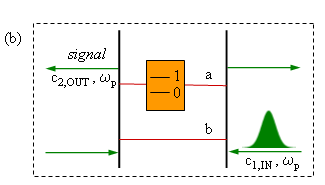}
\caption{\label{fig1}
Schematics of the considered measurement scheme in one- and two-sided cavities. 
(a) The one-sided cavity has two modes ($a$ and $b$, with frequencies $\Omega_a$ and $\Omega_b$) and is pumped at the frequency $\omega_p$. The qubit is coupled to mode $a$ in the dispersive regime, and induces a state-dependent shift $\pm \chi$ in its frequency. The effect of such shift on the phase of the reflected field is measured by a homodyne detector. (b) In the case of a two-sided cavity, with the two spectrally close modes $a$ and $b$, the signal is represented by the transmitted field. The discussed results can be applied to the case of any other observable of interest (besides the qubit state) whose value is correlated to the frequency of mode $a$.
}
\end{center}
\end{figure}

The maximal dependence of the signal on the qubit state is obtained in the (dispersive) strong coupling regime, where the qubit-induced frequency shift is at least of the same order of the cavity mode decay rate. However, the achievement of such strong coupling regime might be technologically challenging, or increase the probability of a measurement back-action. The possible ways of improving the efficiency of the read out process, without compromising the storage and coherent manipulation of the qubit, include a suitable qubit design \cite{Mallet09}, the introduction of a separate mode for read out \cite{Leek10}, and the use of bandpass filters \cite{Jeffrey14}.

 Here we consider an alternative approach, based on the use of an auxiliary resonator mode, spectrally close to the one that is coupled to the qubit. In the last years, multimode generalizations of cavity and circuit QED schemes have been proposed and implemented, with potential applications in quantum technologies \cite{Horak02,McKay15,Sundaresan15,Naik17,Vaidya18}. Pairs of spectrally close modes can be obtained in bidimensional cavities, which have been fabricated with diverse superconducting materials for different technological purposes \cite{Zhu99,Cassinese01,Zhu05}. In particular, the use of a square geometry and the controlled introduction of defects allow one to independently tune the frequencies of the lower modes, as well as the spatial overlap between the respective field distributions \cite{Naji17,Bonizzoni18}. 

\begin{figure}
\begin{center}
\includegraphics[width=0.30\textwidth]{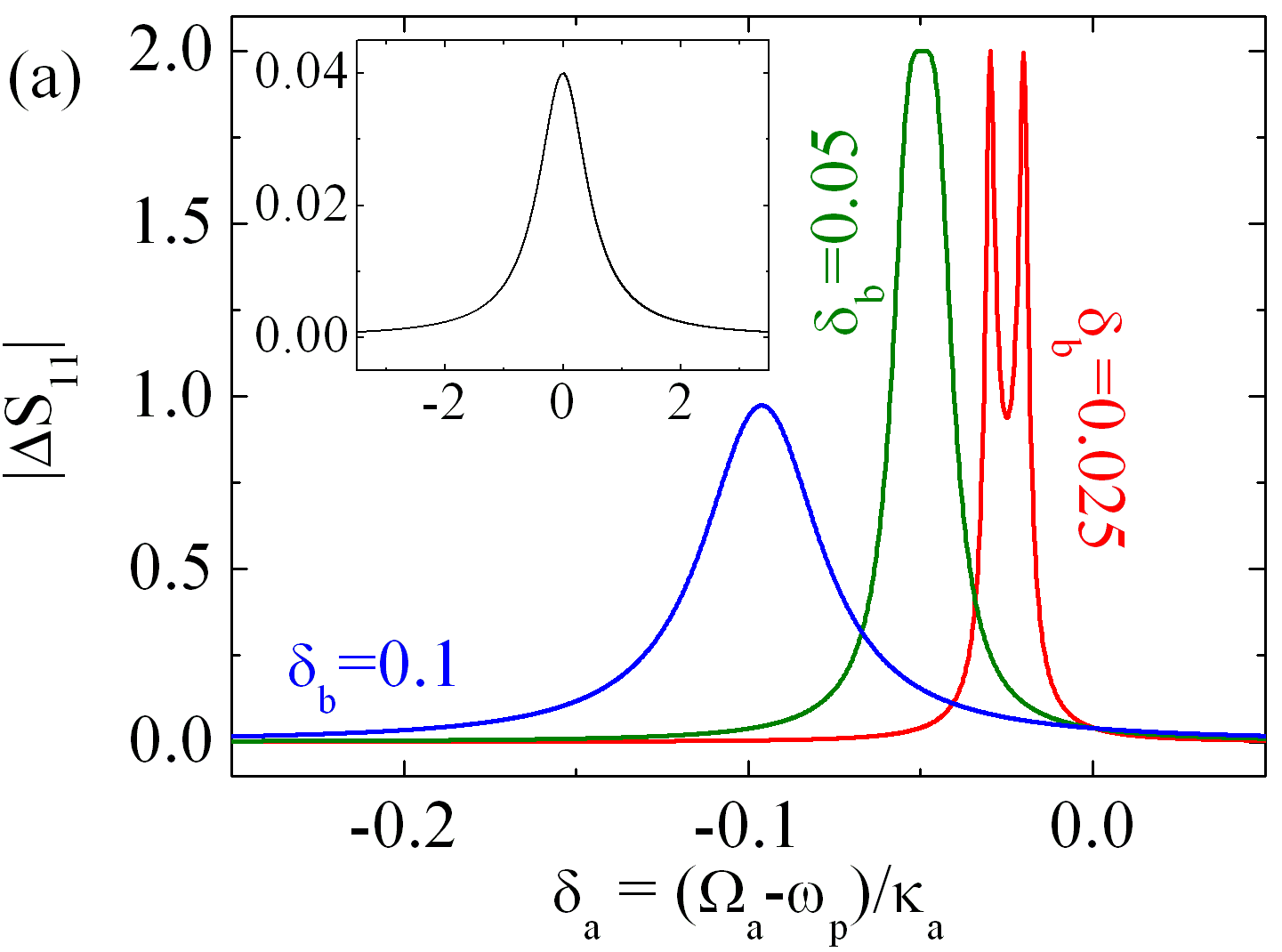}
\includegraphics[width=0.30\textwidth]{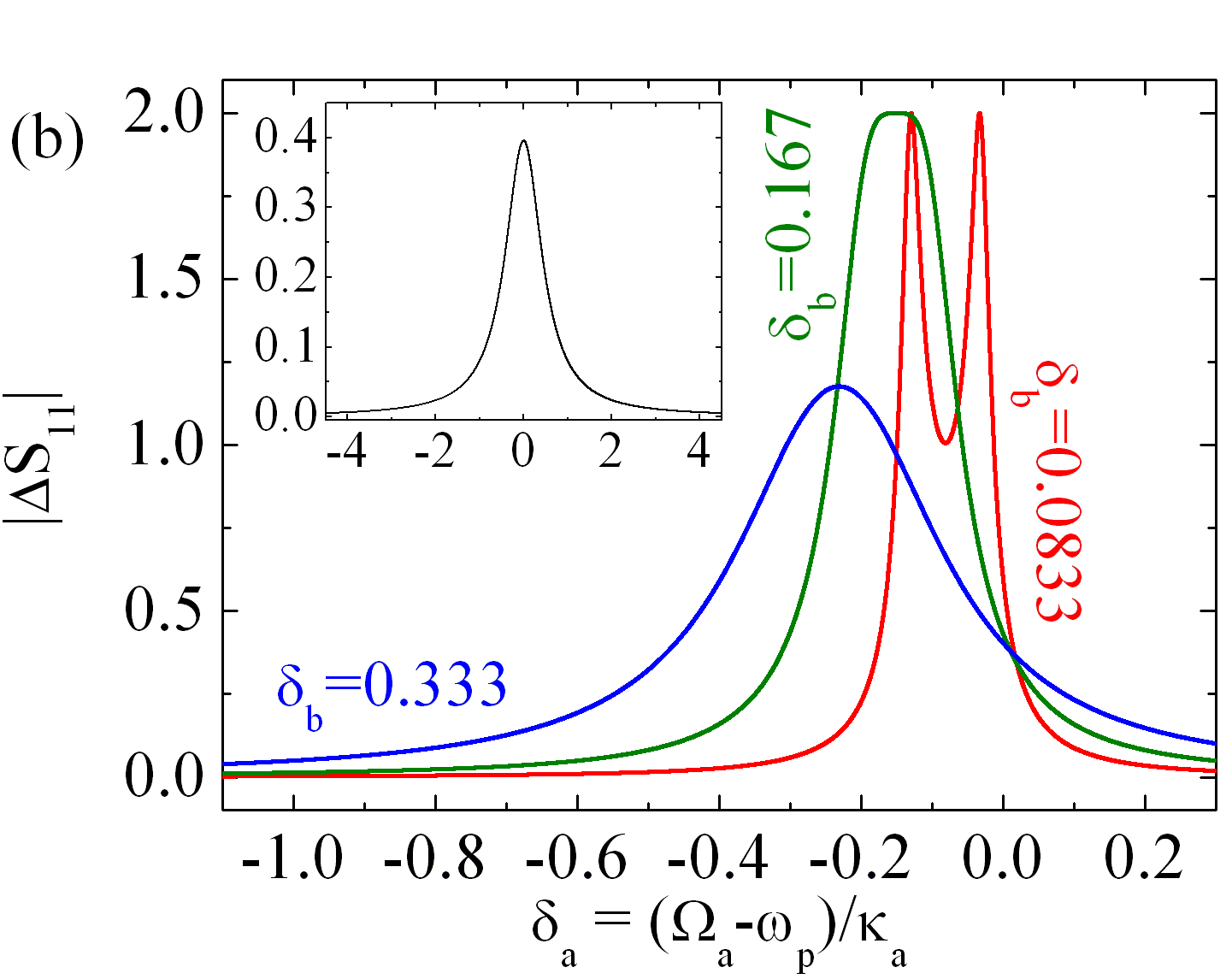}
\includegraphics[width=0.30\textwidth]{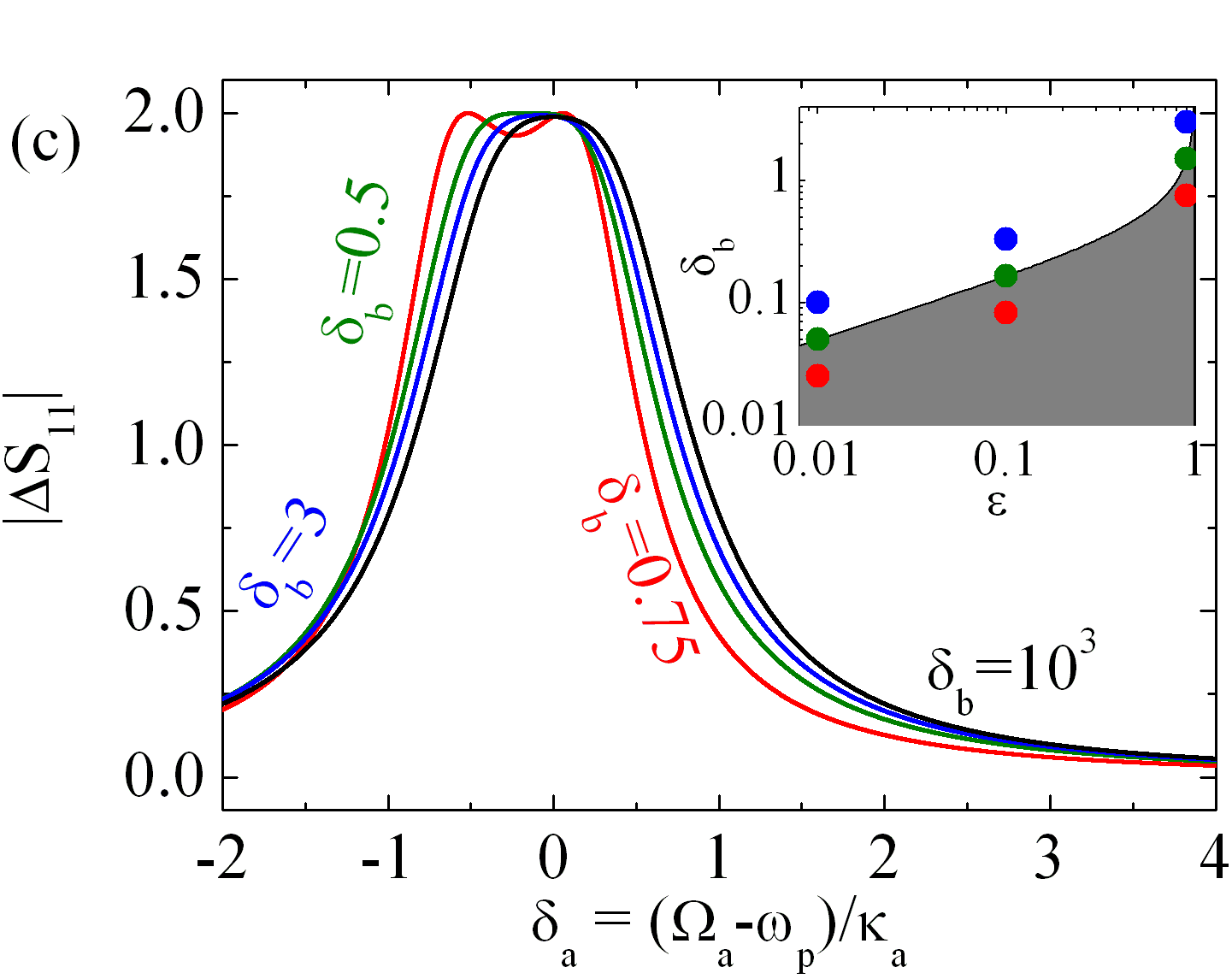}
\caption{\label{fig2}
Difference between the reflection coefficients $S_{11}$ of a one-sided cavity corresponding to the qubit states 0 and 1. The quantity $|\Delta S_{11}|$ is plotted as a function of the detuning between the frequency of the cavity mode $a$ ($\Omega_a$) and the excitation frequency ($\omega_p$), normalized to the mode decay rate ($\kappa_a$). Different panels correspond to different values of the qubit-state dependent frequency shift, normalized to $\kappa_a$: (a) $\epsilon=0.01$, (b) $\epsilon=0.1$, (c) $\epsilon=0.9$. The black curves in panel (c) and in the insets of panels (a,b) are obtained for $\delta_b=10^3$, and can be identified with the case of the single-mode cavity. The inset of panel (c) shows the values of $(\epsilon,\delta_b)$ for which one can have a maximal dependence of the signal on the qubit state ($|\Delta S_{11}|=2$, gray area). The colored dots correspond to the curves displayed in the three panels. 
}
\end{center}
\end{figure}

In the present measurement scheme, a resonator mode (hereafter $b$) is spectrally close to the fundamental one ($a$), which is coupled to the qubit (Fig. \ref{fig1}). The qubit modifies the frequency of mode $a$ in a state-dependent fashion, while $b$ is coupled to the same external field modes as $a$, but not to the qubit. It thus acts as an auxiliary mode, whose role in the scheme is that of enhancing the sensitivity of the measured signal to the small frequency shifts of mode $a$. 
In the case of a one-sided resonator, the signal is identified with the reflected field, whose phase shift is measured by a homodyne detector [panel (a)]. In the case of a two-sided cavity [panel (b)], the signal can be identified with the transmitted field, which generally depends on the qubit state both in phase and amplitude. We note that such scheme can be applied to any case where the observable of interest, here identified with the state of a qubit, is some quantity correlated to the frequency of a cavity mode.

The improvements introduced by the presence of the auxiliary mode are quantified in terms of two figures of merit, expressed as a function of the ratio ($\epsilon$) between the qubit-state dependent frequency shift of mode $a$ ($\xi$) and the mode decay rate ($\kappa_a$). The first figure of merit is represented by the reflection ($S_{11}$) or transmission ($S_{21}$) coefficient and, more specifically, by the difference ($\Delta S_{ij}$) between the values of such coefficient corresponding to small changes in the frequency of mode $a$. An increased sensitivity allowed by the mode $b$ is identified with a large, possibly a maximal, $|\Delta S_{ij}|$ obtained with a given and small epsilon, which would result in a poor dependence of the signal on the qubit state in the absence of the auxiliary mode. The quantity $\Delta S_{ij}$ allows us to characterize the measurement in the stationary regime, and thus to analyze the case of the qubit-state readout, when its lifetime is longer than that of the transient regime. The second figure of merit we consider is the signal-to-noise ratio (SNR), which also accounts for the transient regime and allows one to evaluate the role played by the duration of the measurement. The speed-up allows by the presence of the auxiliary mode can be appreciated by comparing the value of the SNR obtained for a given duration in the two cases.

The paper is organized as follows. Section \ref{SecII} is devoted to the analysis of the measurements in the stationary regime, performed in terms of the reflection and transmission coefficients (respectively for the one- and two-sided cavities).
In Sec. \ref{SecIII} we include the effect of the transient regime and the role played by the measurement duration by computing the signal-to-noise ratio. Finally, we draw the conclusions in Sec. \ref{SecIV}.

\section{Transmission and reflection coefficients}\label{SecII}

We start by considering the reflection coefficient that couples the input and the output modes of one-sided cavities. We seek the conditions under which small variations in the frequency of the fundamental resonator mode $a$ result in large (possibly maximal) changes of this coefficient. This would result in a strong dependence of the phase of the reflected field on the qubit state, and thus in its efficient read out by means of the homodyne detection. An analogous analysis is carried out for the transmitted field in the case of a two-sided cavity.  

\subsection{One-sided cavities}

The frequency-dependent coefficient that couples the input and the output (reflected) fields of a one-sided cavity is given by (see the Appendix A):
\begin{eqnarray}\label{eq00}
S_{11}
&=& 
\frac{\delta_a + \delta_b - 2i \delta_a \delta_b}{\delta_a + \delta_b + 2i 
\delta_a \delta_b} .
\end{eqnarray}
Here, $\delta_\xi \equiv (\Omega_\xi - \omega_p)/\kappa_\xi$ ($\xi = a,b$) is the difference between the frequency $\Omega_\xi$ of the cavity mode $\xi$ and the excitation frequency $\omega_p$, normalized to the mode decay rate $\kappa_\xi$.

Due to the coupling with the qubit (Fig. \ref{fig1}), the frequency of the mode $a$ undergoes a state-dependent shift $\pm \chi$, which we express here in units of the mode decay rate as $\epsilon \equiv 2\chi / \kappa_a$. In the following, we consider the difference between the reflection coefficients $S_{11}$ corresponding to the two qubit states, and thus to the opposite frequency shifts $\pm \chi$: 
\begin{equation}
\Delta S_{11} = S_{11} (\delta_a+\epsilon/2) - S_{11} (\delta_a-\epsilon/2) ,
\end{equation}
and, more specifically, the modulus of such difference. We note that $\Delta S_{11}$ depends on the six parameters $\Omega_\xi$, $\kappa_\xi$ (with $\xi =a,b$), $\chi$, and $\omega_p$ only through their three combinations $\delta_\xi$ and $\epsilon$: in the rest of the present Section, the discussion is thus carried out mainly in terms of these adimensional quantities. 

From the above equations, it follows that $|\Delta S_{11}|$ has a theoretical maximum of 2. In fact, the reflection coefficient can be written in terms of the real number $t=2\delta_a\delta_b/(\delta_b + \delta_a)$ as $S_{11} = (1-it)/(1+it)$, and is a complex number with unitary modulus. The largest possible difference between $S_{11}(t)$ and $S_{11}(t')$, for arbitrary $t$ and $t'$, is thus, in modulus, 2, and is obtained for $S_{11}(t')=-S_{11}(t)$, corresponding to $t'=-1/t$. If the values of the resonator parameters satisfy such condition, the reflected field changes sign as the qubit passes from one logical state to the other. In the following, we investigate more systematically the dependence of $|\Delta S_{11}|$ on the parameters $\delta_a$, $\delta_b$, and $\epsilon$. 

\subsubsection{Optimization of $\epsilon$, $\delta_a$ and $\delta_b$}

Let's start by maximizing $|\Delta S_{11}|$ as a function of $\delta_a$, for given values of $\delta_b$ and $\epsilon$. In particular, we assume that the resonator is probed by a field whose detuning from the auxiliary mode $b$ corresponds to a given value of $\delta_b = (\Omega_b-\omega_p)/\kappa_b $. One can have two different scenarios, depending on the value of state-dependent frequency shift $\epsilon$. In fact, for 
\begin{equation}
\epsilon < \epsilon_{th} \equiv 4\delta_b^2/(1+4\delta_b^2), 
\end{equation}
there is one optimal value of the detuning $\delta_a$, given by
\begin{equation}\label{eqP20}
\delta_a^1 = -\frac{\delta_b}{1+4\delta_b^2},
\end{equation}
for which $|\Delta S_{11}^1|$ displays a maximum, corresponding to
\begin{equation}\label{eqrr}
|\Delta S_{11}^1| = \frac{4[4\delta_b^2][\epsilon(4\delta_b^2+1)]}{[4\delta_b^2]^2+[\epsilon(4\delta_b^2+1)]^2} 
= 
\frac{4\epsilon\epsilon_{th}}{\epsilon^2+\epsilon_{th}^2}. 
\end{equation}
Therefore, in the range of values $[0,\epsilon_{th}]$, the difference in reflection corresponding to the two qubit states increases monotonically with $\epsilon$, and achieves the theoretical maximum of 2 for $\epsilon=\epsilon_{th}$. We note that the threshold value $\epsilon_{th}$ can be made arbitrarily small either by reducing the difference between the excitation frequency $\omega_p$ and the frequency $\Omega_b$ of the auxiliary mode, or, equivalently, by increasing the decay rate $\kappa_b$.

For values of $\epsilon$ that are larger than $\epsilon_{th}$, the optimal values of the detuning $\delta_a$ are given by
\begin{equation}\label{eqP06}
\delta_a^{2,3}\!=\!\frac{-2\delta_b\pm [\epsilon^2(1+4\delta_b^2)^2-16\delta_b^4]^{1/2}}{2(1+4\delta_b^2)}
\!=\!\delta_a^{1}\pm\frac{1}{2}(\epsilon^2-\epsilon^2_{th})^{1/2}
\end{equation}
where $|\Delta S_{11}|$ achieves the theoretical maximum of 2. These two maxima are separated by a minimum, positioned at $\delta_a^1$, where $|\Delta S_{11}|=|\Delta S_{11}^1|$ [see Eqs. (\ref{eqP20},\ref{eqrr})].

The equations that apply in the absence of the auxiliary mode are obtained in the limit of a large detuning, {\it i.e.} for $|\delta_b| \gg |\delta_a|$. In such a limit, one has that $S_{11}=(1-2i\delta_a)/(1+2i\delta_a)$ \cite{WallsMilburn} and $\epsilon_{th} = 1$. This implies that a maximal state-dependent change in the transmission coefficient ($|\Delta S_{11}|=2$) can only be obtained if the frequency shift is at least comparable to the linewidth of the cavity mode $a$ ($\chi=\kappa_a/2$ or, equivalently, $\epsilon=1$). For smaller qubit-dependent frequency shifts, {\it i.e.} for $\epsilon<1$, the maximum of $|\Delta S_{11}|$ is obtained by exciting the cavity resonantly ($\delta_a=\delta^1_a=0$) and is given by $|\Delta S^1_{11}|=4\epsilon/(1+\epsilon^2)<2$ [see Eqs. (\ref{eqP20},\ref{eqrr})]. 

The presence of two regimes can also be derived by starting from a given frequency shift $\epsilon$. In this case, one can derive a threshold value $\delta_{b,th}$ for the detuning between the excitation frequency and that of the auxiliary mode. In fact, one can show that $|\Delta S_{11}|$ displays a single maximum for $\delta_a=\delta_a^1$ [Eqs. (\ref{eqP20},\ref{eqrr})] if 
\begin{equation}\label{eqfjf}
|\delta_b| > \delta_{b,th} \equiv \frac{1}{2}\left(\frac{\epsilon}{1-\epsilon}\right)^{1/2} , 
\end{equation}
while it has two maxima in $\delta_a^{2,3}$ [Eq. (\ref{eqP06})], and achieves the value of 2, for $|\delta_b| \le \delta_{b,th}$. Therefore, the smaller the frequency shift $\epsilon$, the closer the excitation frequency has to be to that of the auxiliary mode $b$ in order for the reflected field to display a maximal dependence on the qubit state, corresponding to $|\Delta S_{11}^1|=2$ [see the inset of Fig. \ref{fig2}(c)].
On the other hand, if $\epsilon$ approaches 1, the excitation frequency can be far detuned from that of mode $b$, and the difference between the cases of the single- and two-mode resonators becomes negligible. 

In order to provide some further insight into the dependence of the reflection coefficient on the system parameters, we plot in Fig. \ref{fig2} the values of $|\Delta S_{11}|$ as a function of $\delta_a$, for a few representative values of $\epsilon$ and $\delta_b$. We start by considering the case $\epsilon = 0.01$ [panel (a)], corresponding to a dispersive coupling constant $\chi$ which is 200 times smaller than the decay rate $\kappa_a$ of the cavity mode. If the detuning of mode $b$ exceeds the threshold value (in particular, for $\delta_b=2\delta_{b,th}$, blue curve), the peak corresponding to the maximum of $|\Delta S_{11}|$ as a function of $\delta_a$ is quite broad, but remains below the theoretical maximum of 2. For $\delta_b=\delta_{b,th}/2$ (red), there are two values of $\delta_a$ for which $|\Delta S_{11}|=2$. At the boundary between the two above regimes ($\delta_b=\delta_{b,th}$, green), $|\Delta S_{11}|$ displays a single maximum, corresponding to the theoretical maximum of 2. In the inset, we plot for a comparison the case of a single-mode cavity, obtained by introducing a large detuning $\delta_b$. In this case, the maximum is two orders of magnitude smaller than that achievable with a two-mode cavity. The difference between the curves corresponding to the four values of $\delta_b/\delta_{b,th}$ is reduced for $\epsilon=0.1$ [panel (b)]. The gain with respect to the single-mode cavity (inset), however, is still significant. The four curves tend to overlap for values of the qubit-induced frequency shift that are comparable to the linewidth of the mode $a$ [$\epsilon=0.9$, panel (c)]. Here, there is no significant gain resulting from the presence of the second mode ($b$) spectrally close to the first one ($a$).

\subsubsection{Optimization in terms of mode splitting and excitation frequency}

Let's consider the case where one has a given ratio between the dispersive coupling $\chi$ and the decay rate $\kappa_a$, and wants to tune the mode splitting of the cavity and the excitation frequency in order to maximize the difference between the signal obtained for the two qubit states. In view of the above analysis, the normalized detunings of modes $b$ and $a$ should correspond to (at most) $\delta_{b,th}$ [Eq. (\ref{eqfjf})] and to $\delta_a^1$ [Eq. (\ref{eqP20}), with $\delta_b=\delta_{b,th}$], respectively. The optimal difference between the frequencies of the cavity modes is thus given by:
\begin{eqnarray}
\Omega_a - \Omega_b 
&=& 
\kappa_a \delta^1_a (\delta_b=\delta_{b,th}) - \kappa_b\delta_{b,th} 
\nonumber\\
&=& 
-\frac{\kappa_a}{2} [\epsilon   (1-\epsilon)]^{1/2}
-\frac{\kappa_b}{2} [\epsilon / (1-\epsilon)]^{1/2}
.
\end{eqnarray}
From the same equations, it follows that the optimal excitation frequency $\omega_p$ reads: 
\begin{eqnarray}
\omega_p \!&=&\! \frac{1}{2}[\Omega_a + \Omega_b
 -\kappa_a \delta^1_a(\delta_{b}=\delta_{b,th}) -\kappa_b\delta_{b,th} ]
\nonumber\\
\!&=&\!
\frac{1}{2}(\Omega_a \!+\! \Omega_b) \!+\!
 \frac{\kappa_a}{4} [\epsilon   (1\!-\!\epsilon)]^{1/2}
\!-\!\frac{\kappa_b}{4} [\epsilon / (1\!-\!\epsilon)]^{1/2} .
\end{eqnarray}
Therefore, one obtains a resonant excitation and quasi-degenerate cavity modes ($\omega_p\simeq\Omega_a\simeq\Omega_b$) in the limit $\epsilon \ll 1$ (weak dispersive coupling). In the opposite limit, where $\epsilon$ approaches 1 (strong dispersive coupling) from below, the excitation should be resonant with the fundamental mode ($\omega_p\simeq\Omega_a$), while the auxiliary mode can be far detuned: this in practice corresponds to a single-mode resonator. For intermediate values of $\epsilon$,  mode $a$ should be spectrally separated from $b$, and slightly detuned from the excitation frequency $\omega_p$ (at most $\kappa_a/4$, for $\epsilon=1/2$).

\subsection{Two-sided cavities}

In the case of a two-sided cavity, the signal is represented by the transmitted field. The frequency-dependent coefficient that couples the input modes on side 1 with the output modes on side 2 is given by:
\begin{equation}\label{eqP01}
S_{21} 
= 
\frac{\delta_b + \delta_a}{\delta_b + \delta_a + i 
\delta_a \delta_b} 
\end{equation}
where, as in the previous case,
$\delta_\xi \equiv (\Omega_\xi - \omega)/\kappa_\xi$ ($\xi = a,b$).
The above expression can be closely related to the reflection coefficient obtained for the one-sided cavity. In fact, one can show that:
$
S_{11}(\delta_a,\delta_b) = 2 S_{21}(2\delta_a,2\delta_b)-1
$. 

The dispersive coupling of the qubit to the mode $a$ gives rise to a state-dependent frequency shift $\pm \chi $, with $\chi=\epsilon\kappa_a/2$. This results in a difference in the transmission corresponding to the two qubit states:
\begin{eqnarray}\label{eq01}
\Delta S_{21}
= S_{21} (\delta_a+\epsilon/2) - S_{21} (\delta_a-\epsilon/2) .
\end{eqnarray} 

It is easily seen that such quantity has a theoretical maximum of 1. In fact, in view of Eq. (\ref{eqP01}), one can always write: $S_{21} = (1+it)^{-1}$, with $t=\delta_a\delta_b/(\delta_b + \delta_a)$ a real number. As a result, the representation of $S_{21}$ in the plane defined by $x={\rm Real} (S_{21})$ and $y={\rm Imag} (S_{21})$ corresponds to a circle or radius 1/2 centered in $(0,1/2)$. This implies that the largest possible difference between $S_{21}(t)$ and $S_{21}(t')$ is, in modulus, 1 and is obtained if the $t'=-1/t$. We note that, unlike the case of one-sided cavity, such maximal difference generally corresponds to a change in both the phase and amplitude of the transmitted field.

The dependence of $|\Delta S_{21}|$ on the relevant physical parameters parallels that of $|\Delta S_{11}|$ in the case of the one sided cavity. In fact, for a given $\delta_b$, one can identify two regimes, corresponding to different ranges of values of the qubit-dependent frequency shift $\epsilon$. In particular, for 
\begin{equation}
\epsilon < \epsilon_{th} \equiv 2\delta_b^2/(1+\delta_b^2), 
\end{equation}
there is one maximum at
\begin{equation}\label{eq02}
\delta_a^1 = - \frac{\delta_b}{1+\delta_b^2} ,
\end{equation}
where the modulus of $\Delta S_{21}$ takes the value
\begin{equation}\label{eq05}
|\Delta S_{21}^1| = \frac{2[2\delta_b^2][\epsilon(\delta_b^2+1)]}{[2\delta_b^2]^2+[\epsilon(\delta_b^2+1)]^2} 
= 
\frac{2\epsilon\epsilon_{th}}{\epsilon^2+\epsilon_{th}^2}. 
\end{equation}
This modulus increases with $\epsilon$, ranging from 0 (for $\epsilon=0$) to 1 (for $\epsilon = \epsilon_{th}$). As in the case of a one-side cavity, we note that such threshold value $\epsilon_{th}$, and thus the qubit-induced energy shift required for a maximal $|\Delta S_{21}|$, can be made arbitrarily small by reducing the normalized detuning $\delta_b$ of the auxiliary mode.
For larger frequency shifts, $\epsilon > \epsilon_{th}$, there are two maxima at the normalized detunings of the fundamental mode given by:
\begin{equation}\label{eq03}
\delta_a^{2,3} 
= 
\frac{-2\delta_b\pm[\epsilon^2(\delta_b^2+1)^2-4\delta_b^4]^{1/2}}{2(\delta_b^2+1)} 
=
\delta_a^1 \pm \frac{1}{2} (\epsilon^2-\epsilon_{th}^2)^{1/2} ,
\end{equation}
where 
$|\Delta S_{21}|$ achieves the theoretical maximum of 1, while Eqs. (\ref{eq02}-\ref{eq05}) define a minimum. 

For a given value of the frequency renormalization $\epsilon$, the two regimes described above are obtained for values of $\delta_b$ that are respectively larger or smaller than the threshold value
\begin{equation} 
|\delta_b| > \delta_{b,th} \equiv [\epsilon/(2-\epsilon)]^{1/2} .
\end{equation}
Therefore, for any given $\epsilon < 2$, there is a maximal detuning with respect to the mode $b$ which is compatible with the achievement of the theoretical maximum for $|\Delta S_{21}|$.

The equations for the single-mode resonator can be obtained from the above ones in the limit $|\delta_b| \gg |\delta_a|$. In such a limit, one has that $S_{21}=1/(1+i\delta_a)$  \cite{WallsMilburn} and $\epsilon_{th}=2$. Therefore, the achievement of the theoretical maximum for $|\Delta S_{21}|$ requires a frequency shift comparable to the linewidth of the cavity mode $a$ ($\chi=\kappa_a$ or, equivalently, $\epsilon=2$). A visualization of the above results can be given by the plots of $|\Delta S_{11}|$ reported in Fig. \ref{fig2}, considered that
\begin{equation}
\Delta S_{11} ( \delta_a, \delta_b, \epsilon) 
= 
 2\Delta S_{21} (2\delta_a,2\delta_b,2\epsilon) ,
\end{equation}
where the two coefficients $S_{11}$ and $S_{21}$ refer to the one- and the two-sided cavities, respectively. 

\begin{figure}
\begin{center}
\includegraphics[width=0.45\textwidth]{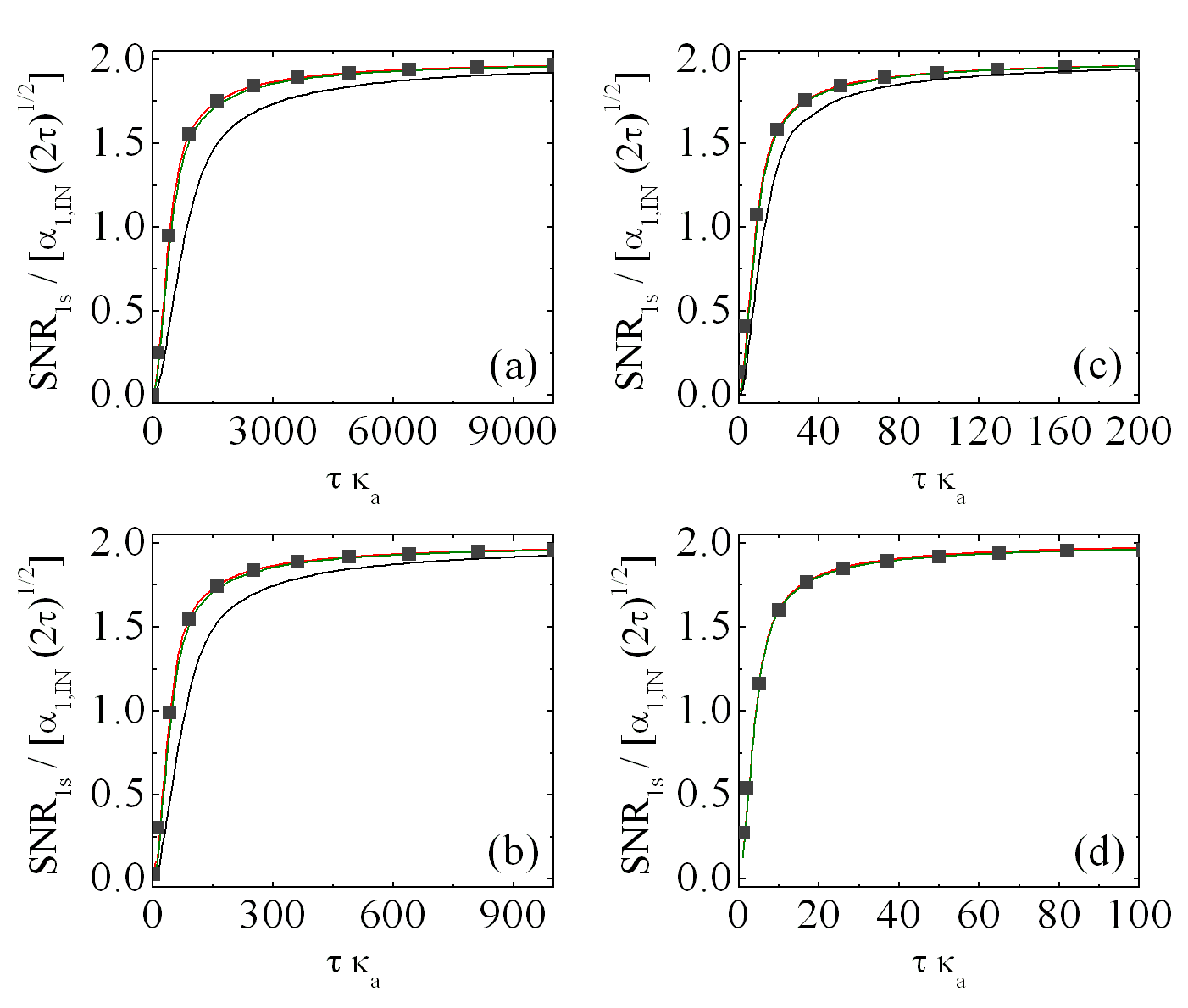}
\caption{\label{fig3}
Signal-to-noise ratio of the one-side cavity obtained for different values of the qubit-dependent shift $\epsilon=2\chi$ (being $\kappa_a=1$): 0.01 (a), 0.1 (b), 0.5 (c), 1.0 (d). The curves in the four panels are obtained for $\delta_b=\delta_{b,th}$, $\delta_a=\delta_a^1$. The value of $\kappa_b$ is either set equal to $\kappa_a$ (black curves), or to $10\kappa_a$ (green), or optimized numerically (red). Gray squares correspond to an unconstrained numerical optimization of the parameters $\Delta_\xi$ and $\kappa_b$, while $\kappa_a=1$ specifies the units.} 
\end{center}
\end{figure}

\section{Signal-to-noise ratio}\label{SecIII}

The reflection and transmission coefficients represent the ratio between the input and the output modes in the case where the system is excited at a given frequency and in the stationary regime. In the case of a measurement of finite duration, where the transient regime plays a non-negligible role, the relevant figure of merit is represented by the signal-to-noise ratio (SNR), which we compute hereafter for the cases of the one- and two-sided cavities. 

\subsection{One-sided cavity}

In order to compute the SNR, we first solve the Heisenberg equations for the annihilation operators corresponding to the two cavity modes. In the case of the one-sided cavity, these are given by:
\begin{equation}\label{equuu}
\frac{d}{dt} 
\left( 
\begin{array}{c}
a 
\\
b
\end{array}
\right)
= 
\mathcal{M}_{1s}
\left( 
\begin{array}{c}
a 
\\
b
\end{array}
\right)
+
\left( 
\begin{array}{c}
\sqrt{\kappa_a} 
\\
\sqrt{\kappa_b}
\end{array}
\right)
c_{1,IN} , 
\end{equation}
where the matrix $\mathcal{M}_{1s}$ reads:
\begin{equation}\label{eqvvv}
\mathcal{M}_{1s} = 
\left( 
\begin{array}{cc}
-i\Omega_a-\frac{1}{2}\kappa_a &
-\frac{1}{2}\sqrt{\kappa_a\kappa_b} \\
-\frac{1}{2}\sqrt{\kappa_a\kappa_b} &
-i\Omega_b-\frac{1}{2}\kappa_b
\end{array}
\right) .
\end{equation}
The Heisenberg equations have to be combined with the boundary condition, which relates the cavity and the external fields:
\begin{equation}\label{eqbcbc}
c_{1,OUT}+c_{1,IN} = a \sqrt{\kappa_a} + b \sqrt{\kappa_b} .
\end{equation}

By solving the above system of equations, and assuming that the input field is given by a coherent state at the frequency $\omega_p$, one obtains the following expression:
\begin{equation}\label{eqP07}
\langle c_{1,OUT} (t) \rangle
=
\langle c_{1,IN} (t) \rangle
\left[ 
\sum_{k=1}^2 A_k^{1s} \left(1-e^{\lambda_kt}\right) -1
\right],
\end{equation}
where $\lambda_k$ are the eigenvalues of the matrix $(\mathcal{M}_{1s} + \omega_p \mathcal{I})$, with $\mathcal{I}$ the identity operator. The expression of the coefficients $A_k^{1s}$, along with further details on the derivation of the above equations, are given in the Appendices, where we also show that the sum of the coefficients $A_k^{1s}$ coincides with the matrix element that couples the input and the output modes:
\begin{equation}\label{eqrrrs}
S_{11} = A_1^{1s} + A_2^{1s} -1.
\end{equation}
This is consistent with the fact that the eigenvalues $\lambda_k$ are complex numbers with negative real components. The exponential terms in Eq. (\ref{eqP07}) thus decay in time and vanish for sufficiently large values of $t$, such that, in the stationary state, one recovers the standard input-output relation in terms of the reflection coefficient $S_{11}$.

In order to compute the SNR, we derive the expectation value of a suitable quadrature $Y$, defined by the angle $\alpha$, which is given by:
\begin{eqnarray}\label{eq009}
\langle Y_l^{1s} (t) \rangle
\!=\!
2|\alpha_{1,IN}| 
{\rm Real} 
\left\{
\!
e^{i\alpha}
\!
\left[ 
\sum_{k=1}^2
A_{k,l}^{1s} \left( 1 \!-\! e^{\lambda_{k,l}t} \right)
\!-\!1\!
\right]\!
\right\}\! ,
\end{eqnarray}
with $\alpha_{1,IN}=\langle c_{1,IN} (t) \rangle e^{i\omega_pt}$ and $l=0,1$ the two qubit states. 
Then, we integrate such quantity over the time interval $[0,\tau]$, with $\tau$ the measurement duration, and normalize such integral to the noise:
\begin{eqnarray}\label{eq09}
SNR_{1s}
\!\!&=&\!\! 
\frac{1}{\sqrt{2\tau}} 
\left| 
\int^\tau_0 dt\, [\langle Y^{1s}_0 (t)\rangle - \langle Y^{1s}_1 (t) \rangle ] 
\right|
= |\alpha_{1,IN}| \sqrt{2\tau} 
\nonumber\\
\!\!&\times&\!\!\!
\left| 
{\rm Real}\!
\left[
e^{i\alpha}\!\sum_{k=1}^2\sum_{l=0}^1  
(-1)^l\!
A_{k,l}^{1s} \!\left( \!\!
1 \!\!+\! \frac{1\!-\!e^{\lambda_{k,l}\tau}}{\lambda_{k,l}\tau}  
\right)\!
\right]
\right|.
\end{eqnarray}
The optimal value of the angle $\alpha$ can be derived from the reflection coefficient $S_{11}$, and in fact coincides with the opposite of its argument: 
$
\Delta S_{11}
=
e^{-i\alpha}|\Delta S_{11}|
$.

In order to gain some intuition on the dependence of the SNR on the relevant physical parameters, we plot in Fig. \ref{fig3} a few representative cases. In the plots, $SNR_{1s}$ is normalized to $\alpha_{1,IN}\sqrt{2\tau}$, in order to single out the dependence on the parameters that characterize the cavity modes. We start by considering the case where $\chi$ is two orders of magnitude smaller than $\kappa_a$ [panel (a)], for $\delta_b=\delta_{b,th}$ and $\delta_a=\delta_{a}^1$, corresponding to the positions of the maximum in $|\Delta S_{11}|$ [see Eq. (\ref{eqP20},\ref{eqfjf})]. We note that, while $\Delta S_{11}$ depends on $\kappa_a$ and $\Delta_a$ ($\kappa_b$ and $\Delta_b$) only through their ratio, $SNR_{1s}$ also depends on the absolute values of such parameters. In fact, the signal-to-noise ratio obtained by setting $\kappa_b=\kappa_a$ (black curves) can be improved by numerically optimizing $\kappa_b$ (red), or simply be enhancing its value by one order of magnitude (green), while leaving the ratio $\Delta_b/\kappa_b$ unaffected. An unconstrained numerical optimization of the parameters $\Delta_\xi$ and $\kappa_b$ (with $\kappa_a=1$, grey squares) doesn't lead to a further enhancement of the signal-to-noise ratio. A similar behavior is obtained in the case where $\chi$ is one order of magnitude smaller than $\kappa_a$ [panel (b)]. The main difference with respect to the previous case is represented by the fact that the dynamics is faster, {\it i.e.} the curve approaches the asymptotic value for smaller values of $\tau$. As the value of $\chi$ approaches that of $\kappa_a$ [panels (c,d)], the improvements in the SNR produced by the optimization of $\kappa_b$ (with respect to the case $\kappa_b=\kappa_a$) become less significant. The same applies to the difference with respect to the asymptotic values ($\tau\rightarrow\infty$) that are obtained in the single-mode case ($|\delta_b| \gg |\delta_a|$), which are approximately given by the following numbers: 0.04, 0.396, 0.625, 2, for $\epsilon = 0.01, 0.1, 0.5, 1$, respectively. 

From the above analysis, one can draw the following conclusions. First, the optimization of $\delta_a$ and $\delta_b$ that is based on the reflection coefficient $S_{11}$ (limit of infinite $\tau\kappa_a$), also leads to a maximal SNR for finite measurement durations. Second, for optimized values of the adimensional parameters $\epsilon$, $\delta_a$ and $\delta_b$, higher values of the SNR are obtained for a leaky auxiliary mode (large $\kappa_b/\kappa_a$). Third, while in the stationary limit one can maximize the SNR for arbitrarily small values of $\epsilon$ by suitably adjusting the detunings $\delta_a$ and $\delta_b$, a reduction of $\epsilon$ implies a longer duration of the transient regime. In fact, as emerges by the comparison of the four panels in Fig. \ref{fig3}, the stationary regime is approached for $\tau\kappa_a\sim 10/\epsilon$. 

\subsection{Two-sided cavity}

In the case of a two-sided cavity, the signal is represented by the transmitted field. Following the same procedure outlined in the previous Subsection, we start by writing the Heisenberg equations for the annihilation operators of the cavity modes:
\begin{equation}
\frac{d}{dt} 
\left( 
\begin{array}{c}
a 
\\
b
\end{array}
\right)
= 
\mathcal{M}_{2s}
\left( 
\begin{array}{c}
a 
\\
b
\end{array}
\right)
+
\left( 
\begin{array}{c}
\sqrt{\kappa_a} 
\\
\sqrt{\kappa_b}
\end{array}
\right)
(c_{1,IN}+c_{2,IN}) , 
\end{equation}
where we assume that the decay rates of each mode at the two sides of the cavity coincide. The matrix $\mathcal{M}_{2s}$ reads:
\begin{equation}
\mathcal{M}_{2s} = 
\left( 
\begin{array}{cc}
-i\Omega_a-\kappa_a &
-\sqrt{\kappa_a\kappa_b} \\
-\sqrt{\kappa_a\kappa_b} &
-i\Omega_b-\kappa_b
\end{array}
\right) .
\end{equation}
The boundary condition relating the cavity and the external fields is given by:
\begin{equation}
c_{k,OUT}+c_{k,IN} = a \sqrt{\kappa_a} + b \sqrt{\kappa_b} ,
\end{equation}
where $k=1,2$ denotes the side of the cavity.

If the input field is given by a coherent state with frequency $\omega_p$, one obtains the following relation between the input field on side 1 and the output field on side 2:
\begin{equation}\label{eq07}
\langle c_{2,OUT} (t) \rangle
=
\langle c_{1,IN} (t) \rangle
\left[ 
\sum_{k=1}^2 A_k^{2s} \left(1-e^{\mu_kt}\right)
\right],
\end{equation}
where $\mu_k$ are the eigenvalues of the matrix $(\mathcal{M}_{2s}+\omega_p\mathcal{I})$. From a comparison between $\mathcal{M}_{1s}$ and $\mathcal{M}_{2s}$, it emerges that
$
\mu_k(2\Delta_a,2\Delta_b,\kappa_a,\kappa_b) 
= 
2\mu_k(\Delta_a,\Delta_b,\kappa_a/2,\kappa_b/2) 
= 
2\lambda_k(\Delta_a,\Delta_b,\kappa_a,\kappa_n) 
$.
In analogy with what shown in the case of the one-sided cavity, the sum of the coefficients $A_k^{2s}$ coincides with the transmission coefficient:
\begin{equation}
S_{21} = A_1^{2s} + A_2^{2s} ,
\end{equation}
consistently with the intuition that $S_{21}$ should represent the constant of proportionality between the output and the input fields in the limit of a large $\tau$ (stationary regime).
From the above relation, and from the analogous one obtained for $S_{11}$ in the case of a one-sided cavity, it also follows that $
A^{1s} (\Delta_a,\Delta_b,\kappa_a,\kappa_b) 
= 
2A^{2s}(2\Delta_a,2\Delta_b,\kappa_a,\kappa_b) 
= 
2A^{2s}(\Delta_a,\Delta_b,\kappa_a/2,\kappa_b/2) 
$. 

Formally, the expressions of $\langle Y^{2s}_l (t) \rangle$ and $SNR_{2s}$ can be obtained respectively from Eq. (\ref{eq009}) and  Eq. (\ref{eq09}), by replacing the pedices and apices $1s$ with $2s$, the eigenvalues $\lambda_{k,l}$ with $\mu_{k,l}$, and $\alpha$ with the $-$Arg$(\Delta S_{21})$. As the relation between the equations concerning these two cases suggest, the conclusions that have been drawn in the previous Subsection for the field reflected by a one-sided resonator also apply to the field transmitted by the two-sided resonator.

\section{Conclusions}\label{SecIV}

In conclusion, we have shown that the presence of a second cavity mode $b$, spectrally close to the first one $a$, can be exploited in the dispersive readout of the qubit state. In particular, such auxiliary mode in principle allows one to achieve the maximal contrast between the signal corresponding to the two qubit states, for arbitrarily small ratios between the dispersive coupling and the decay rate of mode $a$. This is in stark contrast with the case of a single cavity mode, where such maximal contrast requires the achievement of the dispersive strong coupling regime. The same considerations apply to any case where some quantity of interest is correlated to a frequency shift of the resonator mode. 

The above analysis is performed first in terms of the reflection and transmission coefficients, which represent the relevant figures of merit for the stationary regime. The expressions of such coefficients allow us to derive the optimal working conditions in the form of simple analytical relations between the detuning of the excitation frequency with respect to both modes, the qubit-state dependent frequency shift, and of the mode decay rates. The transient regime is investigated by means of the signal-to-noise ratio. We show that the optimal working conditions derived for the stationary regime also maximize the signal-to-noise ratio at finite measurement durations, if combined with the requirement that the auxiliary mode $b$ has a decay rate much larger than that of the mode $a$.

\appendix

\section{SCATTERING MATRIX}

As a preliminary step, we investigate the eigenvectors of the matrix $\mathcal{M}_{1s}$. This corresponds, up to an additive term proportional to the identity, to a matrix of the form
\begin{equation}
A = 
\left( 
\begin{array}{cc}
z &
x \\
x &
-z
\end{array}
\right) ,
\end{equation}
where $x=-(\kappa_a\kappa_b)^{1/2}$ and $z=i(\Omega_b-\Omega_a)/2+(\kappa_b-\kappa_a)/2$. 
The eigenvalues of $A$ are given by the complex numbers $\bar{\lambda}_{1,2} = \pm (x^2+z^2)^{1/2} $. It is easily seen that: $(i)$ the vectors $\vec{\eta}_k=(\eta_{ka},\eta_{kb})=(x,\bar{\lambda}_k-z)/[2\bar{\lambda}_k(\bar{\lambda}_k-z)]^{1/2}$ are eigenvectors of $A$ with eigenvalues $\bar{\lambda}_k$; $(ii)$ their components satisfy the relations $\sum_{\xi=a,b} \eta_{k\xi} \eta_{k'\xi}=\delta_{k,k'}$, with $\delta_{k,k'}$ the Kronecker delta; $(iii)$ the components of the eigenvectors also satisfy the equations: $\eta_{1a}=-\eta_{2b}$ and $\eta_{1b}=\eta_{2a}$. 

In order to derive the $S$ matrix, that couples the Fourier components of the input and output modes at each given frequency, we start by expressing the cavity modes $a$ and $b$ as a function of the input modes. This is done through the Heisenberg equations in the frequency domain \cite{WallsMilburn}:
\begin{equation}\label{eqfdfd}
\left[ 
\begin{array}{c}
a(\omega) 
\\
b(\omega)
\end{array}
\right]
= 
-(\mathcal{M}_{1s}+\omega\mathcal{I})^{-1}
\left( 
\begin{array}{c}
\sqrt{\kappa_a} 
\\
\sqrt{\kappa_b}
\end{array}
\right)
c_{1,IN} (\omega).
\end{equation}
In view of the above preliminary results, the inverse of the matrix $(\mathcal{M}_{1s}+\omega\mathcal{I})$ can be expressed in the following form:
\begin{equation}
(\mathcal{M}_{1s}+\omega\mathcal{I})^{-1} = 
\sum_{k=1,2}
\frac{1}{\lambda_k}
\left( 
\begin{array}{cc}
\eta_{ka}^2 &
\eta_{ka}\eta_{kb} \\
\eta_{ka}\eta_{kb} &
\eta_{kb}^2
\end{array}
\right) ,
\end{equation}
where $\lambda_k=\bar{\lambda}_k-i(\Omega_a+\Omega_b)/2-(\kappa_a+\kappa_b)/2+\omega$ are the eigenvalues of $(\mathcal{M}_{1s}+\omega\mathcal{I})$.
Combining the resulting expressions for the operators $a(\omega)$ and $b(\omega)$ with the boundary conditions [Eq. \ref{eqbcbc}], one finally obtains:
\begin{equation}\label{eqttt}
S_{11} = \frac{c_{1,OUT}(\omega)}{c_{1,IN}(\omega)} = 
-\sum_{k=1,2} \frac{1}{\lambda_k} \sum_{\xi=a,b} \eta_{k\xi}\eta_{k\xi'} \sqrt{\kappa_\xi \kappa_{\xi'}}  
-1.
\end{equation}
This equation, combined with Eq. \ref{app03} below and with the above property $(iii)$, leads to Eq. \ref{eqrrrs}. 

The expression of $S_{11}$ is more easily obtained by solving the system of equations Eq. \ref{eqfdfd}
\begin{equation}
\left(
\begin{array}{c}
\sqrt{\kappa_a} 
\\
\sqrt{\kappa_b}
\end{array}
\right)
\!
c_{1,IN}(\omega)
\!\equiv\!
\left(\! 
\begin{array}{cc}
m_{11} & m_{12}
\\
m_{21} & m_{22}
\end{array}
\!\right)
\left[ 
\begin{array}{c}
a(\omega)
\\
b(\omega)
\end{array}
\right]
\end{equation}
by linear combination [where $m_{ij}$ are the elements of the matrix $-(\mathcal{M}_{1s}+\omega\mathcal{I})$]. This allows one to express the cavity field operators in terms of the input field operator:
\begin{eqnarray}
a(\omega) &=& \frac{m_{22}\sqrt{\kappa_a}-m_{12}\sqrt{\kappa_b}}{m_{11}m_{22}-m_{12}m_{21}} \, c_{1,IN}(\omega)
\\
b(\omega) &=& \frac{-m_{21}\sqrt{\kappa_a}+m_{11}\sqrt{\kappa_b}}{m_{11}m_{22}-m_{12}m_{21}} \, c_{1,IN}(\omega).
\end{eqnarray} 
Finally, by combining these equations with the boundary condition [Eq. (\ref{eqbcbc})], one obtains the expression of the reflection coefficient [Eq. \ref{eq00}]. 

The results concerning the two-sided cavity can be obtained by proceeding along the same lines. 

\section{Signal-to-noise ratio}

In order to derive the signal-to-noise ratio for the case of the one-sided cavity, we start from the system of Heisenberg equations for the cavity-field operators [Eqs. (\ref{equuu},\ref{eqvvv})]. After computing the eigenvalues $\tilde\lambda_k$ of the matrix $\mathcal{M}_{1s}$, we introduce the auxiliary operators $d_1$ and $d_2$ that diagonalize the system of differential equations, reducing it to:
\begin{eqnarray}\label{app01}
\dot{d}_k = \tilde\lambda_k d_k + (\eta_{ka}\sqrt{\kappa_a}+\eta_{kb}\sqrt{\kappa_b}) c_{1,IN},
\end{eqnarray}
where $k=1,2$ and the coefficients $\eta_{k\xi} $ are given in Appendix A.
The auxiliary operators can be expressed in terms of the cavity field operators through the relations
\begin{equation}
d_k = \eta_{ka} a + \eta_{kb} b ,
\end{equation}
while the inverse transformation can be shown to be
\begin{equation}
a = \zeta_{1a} d_1 + \zeta_{2a} d_2 , \ 
b = \zeta_{1b} d_1 + \zeta_{2b} d_2 ,
\end{equation}
where 
$\zeta_{1a}=-\eta_{2b}$,
$\zeta_{2a}= \eta_{1b}$,
$\zeta_{1b}= \eta_{2a}$,
and
$\zeta_{2b}=-\eta_{1a}$.
After integrating Eq. \ref{app01}, one obtains: 
\begin{eqnarray}
d_k (t) \!=\! e^{\lambda_kt} 
[ 
d_k(0) \!+\!\!\!\sum_{\xi = a,b}\eta_{k\xi}\!\sqrt{\kappa_\xi} 
\!\int^t_0 \!\!\!du\, e^{-\lambda_ku} c_{1,IN} (u)
].
\end{eqnarray}

If the system is excited at the frequency $\omega_p$ by means of a coherent input field, then the expectation value of $c_{1,IN}$ is given by: 
$
\langle c_{1,IN} (t) \rangle = \alpha_{1,IN} e^{-i\omega_p t} 
$.
This equation can be plugged into the previous one, in order to derive the expectation value of the operators $d_k$, which reads:
\begin{eqnarray}
\frac{\langle d_k (t) \rangle}{\langle c_{1,IN} (t) \rangle} = 
(\eta_{ka}\sqrt{\kappa_a}+\eta_{kb}\sqrt{\kappa_b}) 
\frac{e^{\lambda_kt}-1}{\lambda_k} 
\equiv D_k(t) ,
\end{eqnarray}
with $\lambda_k=\tilde\lambda_k+i\omega_p$.
Combining the above equation with the boundary conditions [Eq. (\ref{eqbcbc})], one finally obtains:
\begin{equation}
\frac{\langle c_{1,OUT} (t) \rangle}{\langle c_{1,IN} (t) \rangle} = 
\sum_{\xi=a,b} \sum_{k=1,2} \sqrt{\kappa_\xi}\,\zeta_{k\xi}\,D_k (t) - 1.
\end{equation}
From this it follows that the coefficients entering the expression of the signal-to-noise ratio $SNR_{1s}$ are given by:
\begin{equation}\label{app03}
A_k^{1s} = -\frac{1}{\lambda_k} \sum_{\xi,\xi'} \sqrt{\kappa_\xi\kappa_{\xi'}}\,\eta_{k\xi} \zeta_{k\xi'} .
\end{equation}
This equation, combined with the Eq. \ref{eqttt} and with property $(iii)$ in Appendix A, leads to Eq. \ref{eqrrrs}. 

From the equations above, one can also derive the expression of the intracavity fields in terms of the expectation values of the operators $a$ and $b$:
\begin{equation}
\frac{\langle a(t) \rangle}{\langle c_{1,IN} (t) \rangle} \!=\! \sum_{k=1,2} \eta_{ka} D_k(t)
,\ 
\frac{\langle b(t) \rangle}{\langle c_{1,IN} (t) \rangle} \!=\! \sum_{k=1,2} \eta_{kb} D_k(t) .
\end{equation} 

The derivation of the signal-to-noise ratio in the case of the two-sided cavities has been carried out along the same lines. 

\begin{acknowledgments}
We thank C. Bonizzoni, A. Ghirri and M. Affronte for useful discussions.
This work is partially funded by the Air Force Office of Scientific Research grant (contract no FA2386-17-1-4040).

\end{acknowledgments}

{99}

\end{document}